\documentclass[superscriptaddress,aps,pra,twocolumn,showpacs,nofootinbib,longbibliography]{revtex4-1}
\usepackage{amsmath,amssymb,amsthm}
\usepackage{easybmat}
\usepackage[colorlinks=true,citecolor=blue,urlcolor=blue]{hyperref}
\usepackage[pdftex]{graphicx}
\usepackage{times,txfonts}
\usepackage{braket}
\usepackage{color}
\usepackage{natbib}
\newcommand{\be}{\begin{equation}}
\newcommand{\ee}{\end{equation}}
\newcommand{\ba}{\begin{eqnarray}}
\newcommand{\ea}{\end{eqnarray}}
\newcommand{\ketbra}[2]{|#1\rangle \langle #2|}
\newcommand{\tr}{\operatorname{Tr}}

\begin{document}

\title{Dynamics of two central spins immersed in spin baths}
\author{Devvrat Tiwari}
\email{devvrat.1@iitj.ac.in}
\affiliation{Indian Institute of Technology Jodhpur, Rajasthan, India}
\author{Shounak Datta}
\affiliation{S. N. Bose National Centre for Basic Sciences, Block JD, Sector III, Salt Lake, Kolkata 700 098, India}
\author{Samyadeb Bhattacharya} 
\email{samyadeb.b@iiit.ac.in}
\affiliation{Center for Security Theory and Algorithmic Research,
International Institute of Information Technology, Gachibowli, Hyderabad, India}
\author{Subhashish Banerjee}
\email{subhashish@iitj.ac.in }
\affiliation{Indian Institute of Technology Jodhpur, Rajasthan, India}

\begin{abstract}
In this article we derive the exact dynamics of a two qubit (spin $1/2$) system interacting centrally with separate spin baths composed of qubits in thermal state. Further, each spin of a bath is coupled to every other spin of the same bath. The corresponding dynamical map is constructed. It is used to analyse the non-Markovian nature of the two qubit central spin dynamics. We further observe the evolution of quantum correlations like entanglement and discord under the influence of the environmental interaction. Moreover, we demonstrate the comparison between this exact two qubit dynamics and the locally acting central spin model in a spin bath. This work is a stepping stone towards the realization of non-Markovian heat engines and other quantum thermal devices.
\end{abstract}

\maketitle 

\section{Introduction}

In the interactive world of quantum particles, it is almost impossible to create a physical system devoid of any external noisy influence. Quantum systems conducive for information theoretic tasks, i.e. trapped ions \citep{ion}, quantum dots \citep{qdot}, NMR qubits \citep{nmr}, Josephson junctions \citep{joseph} and many more, are subjected to environmental interactions. It is therefore a necessary task to investigate the dynamics of such quantum systems, which are under the influence of environmental interactions. The theory of open quantum systems \citep{breuer,banerjee1}, has found numerous applications, in recent times, in quantum information and its interface with various facets of quantum physics \citep{sb1,gorini,lindblad,caldeira,feynman,weiss,legget,nakajima,zwanzig,chaturvedi,banerjee2,banerjee3,banerjee4,banerjee5,banerjee6,banerjee7,banerjee8, naikoo1, Shrikant-dephasing, Naikoo2}. Over the past few decades, our understanding of such systems has stretched from the limitations of Markovian dynamics to the more intriguing and challenging domain of non-Markovian quantum systems \citep{breuer1,laine,rivas,rivas1,vasile,lu,luo,fanchini,titas,haseli,hall1,hall2,sam1,sam2,sam3,sam4,sam5,sam6,sam7,sam8,sam9,banerjee-brownian, kumar2018enhanced, kumar2018non, utagi2020temporal, Krovi-IsingSpinBath}. Even now, it is a challenging task to construct the reduced dynamics of such a system without the Born-Markov and stationary bath approximations \citep{breuer}. One often associates a deviation from quantum dynamical semigroup evolution of a system to a non-Markovian process~\citep{utagi2020temporal}. In a non-Markovian evolution, the time scales of the system and the environment are often not well separated, which can result in information backflow from the environment to the system~\cite{BLP-measure, rivas}. This generally leads to recurrences of quantum properties which is important for a fundamental understanding of system dynamics. 

It would be pertinent to add here that a number of techniques have been developed in recent times to tackle this problem \citep{devega-alonso}. Thus, for example, there are the embedding methods, such as the pseudo-modes technique wherein the decay of an atom strongly coupled to a reservoir can be studied by considering an enlarged system that includes a set of pseudo-modes \citep{garraway-knight, garraway-dalton}. Another relevant technique is the reaction-coordinate mapping \citep{smith-lambert, garg-Onuchic, hughes-christ}. Further, a numerically exact hierarchical equations of motion (HEOM) method has been developed \citep{tanimura-kubo}.

Quantum baths are generally categorized in two broad classes; (a) Bosonic and (b) spin bath. Archetypal examples of Bosonic baths include Caldeira-Leggett model \citep{caldeira} or spin boson model \citep{legget}. Exact quantum master equations for these type of models are common in the literature \citep{breuer}. On the contrary, in the case of spin baths, we often have to rely on perturbative techniques or time non-local master equations \citep{breuer1,breuer2}. Although, the study of these systems are extremely important in physical systems of paramount importance such as magnetic systems, quantum spin glasses and superconducting systems \citep{breuer1}, the theoretical modelling of such systems is still lacking in many different perspectives, especially in bi-partite or multi-partite systems relevant in various quantum device modelling. In this work, we attempt to lay the bedrock of such a construction from the perspective of modelling various quantum devices. Here we develop an exact reduced dynamics of a two qubit system immersed in spin baths, each of which is interacting centrally with the system of interest. Our model Hamiltonian of the two qubit system is inspired from a model of quantum thermal diode \citep{diode}. In recent times, motivated by the goal of building quantum computers, a lot of effort has been given to develop quantum versions of various thermodynamic and electronic devices like refrigerators and heat engines \citep{refri1,refri2,refri3,refri4}, thermal diodes \citep{diode,diode1}, transistors \citep{trans1,trans2,trans3,trans4}, quantum batteries and so on \citep{other,other1}. Drawing the motivation from the long term goal of realizing experimentally feasible models of quantum thermal devices, we develop the aforementioned exact reduced dynamics of a two qubit spin system. We further analyse various thermodynamic and information theoretic properties of the system undergoing the specific open quantum evolution. We also observe the fundamentally non-Markovian behaviour of the dynamical map. 

The flow of the paper is as follows: In Sec. II we introduce the model. Its reduced dynamics is developed in Sec. III. The corresponding dynamical map is discussed next, along with its operator sum representation. The dynamical map is then put to use in Sec. V. This includes analyzing a witness to identify inherent non-Markovianity in the dynamics, the corresponding local and global dynamical maps and finally the quantum correlations, including entanglement and discord, generated between the two spins by their open system dynamics. This is followed by the Conclusion. 
\section{The Model}

\begin{figure}[h]
\includegraphics[width=1\columnwidth]{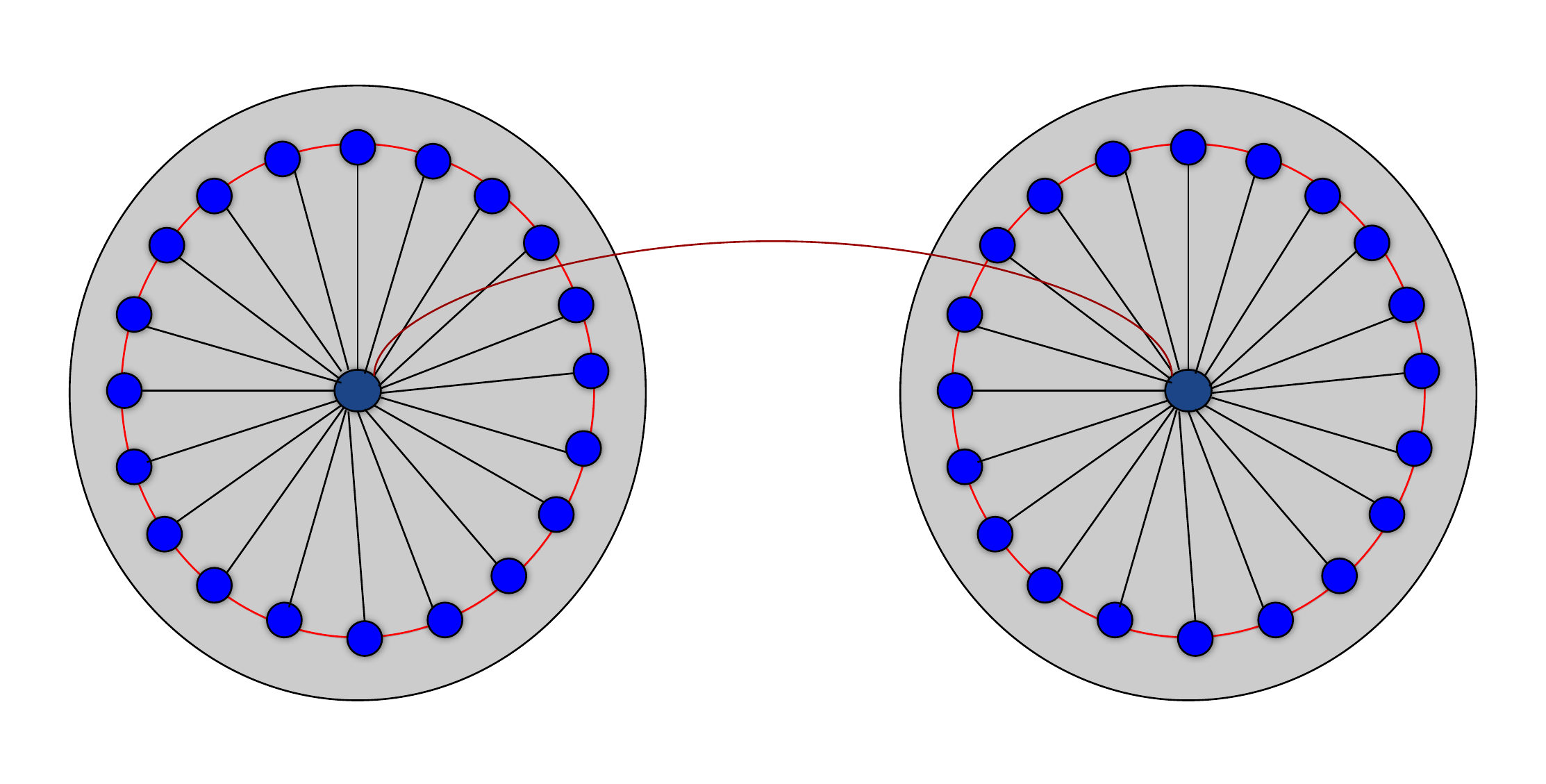}
\caption{(Colour online) Schematic diagram of coupled central spin model where each central spin interacts with individual spin baths.}
\label{figure_1}
\end{figure}

We present a model of two coupled qubits where each qubit is centrally coupled to different thermal spin baths (see Fig. \ref{figure_1}). Further, each spin of a bath is coupled to every other spin of the same bath.

Initially, $\rho_{SB}(0)=\rho_S(0) \otimes \rho_B(0)$, where, $\rho_B(0)=\frac{e^{-H_B/K_B T}}{Z}$ for each central spin immersed in a spin bath. The evolution of the whole system is governed by the following Hamiltonian in our spin-bath model:
\begin{align}\label{1}
H &= H_{S_1} + H_{S_2} + H_{S_1 S_2} + H_{B_1} + H_{B_2} + H_{S_1 B_1} + H_{S_2 B_2}, \nonumber\\
&= \frac{\hslash \omega_1}{2} \sigma_{1z}^0 + \frac{\hslash \omega_2}{2} \sigma_{2z}^0 + \frac{\hslash \delta}{2} (\sigma_{1z}^0 \otimes \sigma_{2z}^0) \nonumber\\
& +\frac{\hslash \omega_a}{2M} \sum_{i=1}^M \Big\{ \frac{1}{2} \sum_{\substack{j=1 \\ j\neq i}}^M (\sigma_{1x}^i \sigma_{1x}^j + \sigma_{1y}^i \sigma_{1y}^j) + \sigma_{1z}^i \Big\}\nonumber\\
& + \frac{\hslash \omega_b}{2N} \sum_{i=1}^N \Big\{ \frac{1}{2} \sum_{\substack{j=1 \\ j\neq i}}^N (\sigma_{2x}^i \sigma_{2x}^j + \sigma_{2y}^i \sigma_{2y}^j) + \sigma_{2z}^i \Big\}\nonumber\\
& + \frac{\hslash \epsilon_1}{2\sqrt{M}} \sum_{i=1}^M (\sigma_{1x}^0 \sigma_{1x}^i + \sigma_{1y}^0 \sigma_{1y}^i) + \frac{\hslash \epsilon_2}{2\sqrt{N}} \sum_{i=1}^N (\sigma_{2x}^0 \sigma_{2x}^i + \sigma_{2y}^0 \sigma_{2y}^i),
\end{align}
where $\sigma_{lk}^i$ or $\sigma_{lk}^j$ (k=x,y,z; $l$=1,2) are Pauli matrices corresponding to $i$-th or $j$-th spin of the $l$-th bath and $\sigma_{lk}^0$ (k=x,y,z; $l$=1,2) are same for $l$-th central spin. $\omega_1$ and $\omega_2$ are the two central spin frequencies and $\delta$ corresponds to their coupling strength. Also, $\omega_a$ and $\omega_b$ are bath frequencies of two spin baths and $\epsilon_l$ ($l$=1,2) are interaction parameters of $l$-th system-bath. $M$ and $N$ are the respective number of atoms in two baths.

Using total spin angular momentum operator, $J_{lk'}=\frac{1}{2}\sum_{i=1}^{M or N} \sigma_{lk'}^i$ ($k'$=x,y,z,+,-; $l$=1 or 2 corresponding to summation upper limit M or N), we may rewrite bath Hamiltonians as,
\begin{eqnarray}
H_{B_1}= \hslash \omega_a (\frac{J_{1+} J_{1-}}{M} - \frac{\openone}{2}), \nonumber\\
H_{B_2}= \hslash \omega_b (\frac{J_{2+} J_{2-}}{N} - \frac{\openone}{2}),
\end{eqnarray}
and system-bath interaction Hamiltonians as,
\begin{eqnarray}
H_{S_1 B_1} =\frac{\hslash \epsilon_1}{\sqrt{M}} (\sigma_{1x}^0 J_{1x} +\sigma_{1y}^0 J_{1y}), \nonumber\\
H_{S_2 B_2} =\frac{\hslash \epsilon_2}{\sqrt{N}} (\sigma_{2x}^0 J_{2x} +\sigma_{2y}^0 J_{2y}).
\end{eqnarray}

Following \cite{single-central-spin}, we then use the Holstein-Primakoff transformation \citep{holstein-primakoff} to redefine collective angular momentum operators as,
\begin{eqnarray}
J_{1+}=\sqrt{M} a^{\dagger}\Big(1-\frac{a^{\dagger}a}{M}\Big)^{1/2}, \nonumber\\
J_{1-}=\sqrt{M} \Big(1-\frac{a^{\dagger}a}{M}\Big)^{1/2} a,
\end{eqnarray}
for first bath and the following for the second bath
\begin{eqnarray}
J_{2+}=\sqrt{N} b^{\dagger}\Big(1-\frac{b^{\dagger}b}{N}\Big)^{1/2}, \nonumber\\
J_{2-}=\sqrt{N} \Big(1-\frac{b^{\dagger}b}{N}\Big)^{1/2} b.
\end{eqnarray}
Here $a$ and $a^{\dagger}$ are the bosonic annihilation and creation operators for the first spin bath having the property, $[a,a^{\dagger}]=1$ and $b$ and $b^{\dagger}$ represents the same for the second spin bath. After this transformation, the bath Hamiltonians appear as,
\begin{eqnarray}
H_{B_1}= \hslash \omega_a \Big\{ a^{\dagger}a \Big(1-\frac{a^{\dagger}a-1}{M}\Big)-\frac{\openone}{2}\Big\}, \nonumber\\
H_{B_2}= \hslash \omega_b \Big\{ b^{\dagger}b \Big(1-\frac{b^{\dagger}b-1}{N}\Big)-\frac{\openone}{2}\Big\},
\end{eqnarray}
and the interaction Hamiltonians of respective spin-baths as,
\begin{eqnarray}
H_{S_1 B_1}=\hslash \epsilon_1 \Big\{ \sigma_{1+}^0 \Big(1-\frac{a^{\dagger}a}{M}\Big)^{1/2} a +\sigma_{1-}^0 a^{\dagger} \Big(1-\frac{a^{\dagger}a}{M}\Big)^{1/2}\Big\}, \nonumber\\
H_{S_2 B_2}=\hslash \epsilon_2 \Big\{ \sigma_{2+}^0 \Big(1-\frac{b^{\dagger}b}{N}\Big)^{1/2} b +\sigma_{2-}^0 b^{\dagger} \Big(1-\frac{b^{\dagger}b}{N}\Big)^{1/2}\Big\}.
\end{eqnarray}

It is also important to mention the limitations and the essential approximations. Here we are taking a central spin half system interacting homogeneously with each of the bath spins with all the characteristic frequencies to be of constant value. We have taken homogeneous interactions for the sake of analytical clarity. This is not an oversimplified assumption, as these kinds of interactions exist in physical situations, of which some examples are quantum spin glasses, superconducting systems and NMR. The more general system with in-homogeneous interaction parameters can only be handled numerically.
A use is then made of the Holstein-Primakoff transformation. A Hamiltonian describing the collective behaviour of $N$ interacting spins can be mapped to a bosonic one employing this transformation, at the expense of having an infinite series in powers of the bosonic creation and annihilation operators. Truncating this series up to quadratic terms allows for obtaining analytic solutions, which become exact in
the limit $N\rightarrow\infty$. In the literature, works on similar spin environments exists which make use of different methods, {\it cf.}~\citep{breuer1}. The reason we choose Holstein-Primakoff transformation here is because of its technical advantage, both from analytical and numerical perspectives. We can work with more number of bath spins with lesser numerical limitations. Homogeneous interactions further enable us to modify the total Hamiltonian into a form of non-linear Jaynes-Cummings model. Other than considering homogeneous interactions, no further assumptions are needed for the type of system-bath model studied here. This opens up an opportunity to study more complex quantum devices and networks.

\section{Reduced dynamics of the two qubit central spin model}
We derive the reduced dynamics of the system of two coupled central spins evolved under the Hamiltonian $H$ along with the two baths by tracing over the bath degrees of freedom.

Consider the evolution of the state $|\psi(0)\rangle=|11\rangle|xy\rangle$, where two central spins are in the excited state $|1\rangle$ and $|x\rangle$ is an arbitrary state for the 1st bath while $|y\rangle$ is a state belonging to the 2nd bath. Global unitary operator corresponding to the evolution under Hamiltonian $H$ can be written as, $U(t)=\exp(-\frac{i H t}{\hslash})$. After the evolution let the state at time $t$ be $|\psi(t)\rangle=\varsigma_1(t)|11\rangle|x'y'\rangle +\varsigma_2(t)|10\rangle|x''y''\rangle +\varsigma_3(t)|01\rangle|x'''y'''\rangle$. We exclude the case when both spins of the system are flipped simultaneously; i.e., $|11\rangle\rightarrow |00\rangle$ transition. The transitions we have considered convey the message sufficiently and hence we choose to exclude the aforementioned transition to avoid unnecessary complications in the calculations.  Now we introduce three time-dependent operators $\hat{A}(t),\hat{B}(t)$ and $\hat{C}(t)$ corresponding to the joint Hilbert space of the two baths such that, $\hat{A}(t)|xy\rangle=\varsigma_1(t)|x'y'\rangle$, $\hat{B}(t)|xy\rangle=\varsigma_2(t)|x''y''\rangle$ and $\hat{C}(t)|xy\rangle=\varsigma_3(t)|x'''y'''\rangle$. Then we have, $|\psi(t)\rangle=\hat{A}(t)|11\rangle|xy\rangle +\hat{B}(t)|10\rangle|xy\rangle +\hat{C}(t)|01\rangle|xy\rangle$.

From the time-dependent Schr\"{o}dinger equation $\frac{d}{dt}|\psi(t)\rangle=-\frac{i H}{\hslash}|\psi(t)\rangle$, we can have,\\ 
\begin{align}
\frac{d \hat{A}(t)}{dt}= &-i\Big[\frac{\omega_1 +\omega_2 +\delta}{2} + \omega_a \Big\{a^{\dagger}a\Big(1-\frac{a^{\dagger}a -1}{M}\Big)-\frac{\openone}{2}\Big\} \nonumber\\
&+ \omega_b \Big\{b^{\dagger}b\Big(1-\frac{b^{\dagger}b -1}{N}\Big)-\frac{\openone}{2}\Big\}\Big]\hat{A}(t) \nonumber\\
&-i\epsilon_2\Big(1-\frac{b^{\dagger}b}{N}\Big)^{1/2}b\hat{B}(t) -i\epsilon_1\Big(1-\frac{a^{\dagger}a}{M}\Big)^{1/2}a\hat{C}(t), \nonumber\\
\frac{d \hat{B}(t)}{dt}= &-i\Big[\frac{\omega_1 -\omega_2 -\delta}{2} + \omega_a \Big\{a^{\dagger}a\Big(1-\frac{a^{\dagger}a -1}{M}\Big)-\frac{\openone}{2}\Big\} \nonumber\\
&+ \omega_b \Big\{b^{\dagger}b\Big(1-\frac{b^{\dagger}b -1}{N}\Big)-\frac{\openone}{2}\Big\}\Big]\hat{B}(t) \nonumber\\
&-i\epsilon_2b^{\dagger}\Big(1-\frac{b^{\dagger}b}{N}\Big)^{1/2}\hat{A}(t), \nonumber\\
\frac{d \hat{C}(t)}{dt}= &-i\Big[\frac{-\omega_1 +\omega_2 -\delta}{2} + \omega_a \Big\{a^{\dagger}a\Big(1-\frac{a^{\dagger}a -1}{M}\Big)-\frac{\openone}{2}\Big\} \nonumber\\
&+ \omega_b \Big\{b^{\dagger}b\Big(1-\frac{b^{\dagger}b -1}{N}\Big)-\frac{\openone}{2}\Big\}\Big]\hat{C}(t) \nonumber\\
&-i\epsilon_1a^{\dagger}\Big(1-\frac{a^{\dagger}a}{M}\Big)^{1/2}\hat{A}(t).
\end{align}
Now we substitute $\hat{A}(t)=\hat{A_1}(t)$, $\hat{B}(t)=b^{\dagger}\hat{B_1}(t)$ and $\hat{C}(t)=a^{\dagger}\hat{C_1}(t)$ and have,
\begin{align}
\frac{d \hat{A_1}(t)}{dt}= &-i\Big[\frac{\omega_1 +\omega_2 +\delta}{2} + \omega_a \Big\{\hat{m}\Big(1-\frac{\hat{m} -1}{M}\Big)-\frac{\openone}{2}\Big\} \nonumber\\
&+ \omega_b \Big\{\hat{n}\Big(1-\frac{\hat{n} -1}{N}\Big)-\frac{\openone}{2}\Big\}\Big]\hat{A_1}(t) \nonumber\\
&-i\epsilon_2\Big(1-\frac{\hat{n}}{N}\Big)^{1/2}(\hat{n}+1)\hat{B_1}(t) \nonumber\\
&-i\epsilon_1\Big(1-\frac{\hat{m}}{M}\Big)^{1/2}(\hat{m}+1)\hat{C_1}(t), \nonumber\\
\frac{d \hat{B_1}(t)}{dt}= &-i\Big[\frac{\omega_1 -\omega_2 -\delta}{2} + \omega_a \Big\{\hat{m}\Big(1-\frac{\hat{m} -1}{M}\Big)-\frac{\openone}{2}\Big\} \nonumber\\
&+ \omega_b \Big\{(\hat{n}+1)\Big(1-\frac{\hat n}{N}\Big)-\frac{\openone}{2}\Big\}\Big]\hat{B_1}(t) \nonumber\\
&-i\epsilon_2\Big(1-\frac{\hat{n}}{N}\Big)^{1/2}\hat{A_1}(t), \nonumber\\
\frac{d \hat{C_1}(t)}{dt}= &-i\Big[\frac{-\omega_1 +\omega_2 -\delta}{2} + \omega_a \Big\{(\hat{m}+1)\Big(1-\frac{\hat m}{M}\Big)-\frac{\openone}{2}\Big\} \nonumber\\
&+ \omega_b \Big\{\hat{n}\Big(1-\frac{\hat{n} -1}{N}\Big)-\frac{\openone}{2}\Big\}\Big]\hat{C_1}(t) \nonumber\\
&-i\epsilon_1\Big(1-\frac{\hat{m}}{M}\Big)^{1/2}\hat{A_1}(t).
\label{one-one}
\end{align}
Here $\hat{m}=a^{\dagger}a$ and $\hat{n}=b^{\dagger}b$ are number operators corresponding to bosonic operators of 1st and 2nd bath, respectively, which have the properties $\hat{m}|m\rangle=m|m\rangle$ and $\hat{n}|n\rangle=n|n\rangle$. Therefore, we define, $\hat{A_1}(t)|mn\rangle=A_1(m,n,t)|mn\rangle$, $\hat{B_1}(t)|mn\rangle=B_1(m,n,t)|mn\rangle$ and $\hat{C_1}(t)|mn\rangle=C_1(m,n,t)|mn\rangle$.

By tracing out bath modes, reduced state of the system ($|11\rangle\langle 11|$) becomes
\begin{align}
\phi(|11\rangle\langle 11|) &=\tr_{B_1 B_2}[|\psi(t)\rangle\langle\psi(t)|] \nonumber\\
&=\frac{1}{Z}\sum_m^M \sum_n^N \Big\{|A_1(m,n,t)|^2 |11\rangle\langle 11| \nonumber\\
&+ (n+1)|B_1(m,n,t)|^2 |10\rangle\langle 10| \nonumber\\
&+ (m+1)|C_1(m,n,t)|^2 |01\rangle\langle 01|\Big\} \nonumber\\
&\times \exp[-\frac{\hslash \omega_a}{K_B T}\{m\Big(1-\frac{m-1}{M}\Big)-\frac{1}{2}\}] \nonumber\\
&\times \exp[-\frac{\hslash \omega_b}{K_B T}\{n\Big(1-\frac{n-1}{N}\Big)-\frac{1}{2}\}]
\end{align}
From the solution of Eq. (\ref{one-one}), one can derive the values of $|A_1(m,n,t)|^2$, $|B_1(m,n,t)|^2$ and $|C_1(m,n,t)|^2$ following the steps as mentioned in the Appendix B. The partition function in the above equation is, $Z=\sum_m^M \sum_n^N \exp[-\frac{\hslash \omega_a}{K_B T}\{m\Big(1-\frac{m-1}{M}\Big)-\frac{1}{2}\}] ~ \exp[-\frac{\hslash \omega_b}{K_B T}\{n\Big(1-\frac{n-1}{N}\Big)-\frac{1}{2}\}]$.

In a similar way, we can derive the evolution of the other elements of the system's reduced state. The details are given in Appendix A.
The reduced state of the system of two central spins after the global unitary evolution of the pair of joint system-bath state can be denoted as
\begin{align}
\rho_{S_1 S_2} &= \tr_{B_1 B_2} \Big[e^{-i H t/\hslash} \rho_{SB}(0) e^{i H t/\hslash} \Big] \nonumber\\
&= \begin{pmatrix}
\rho_{11}(t) & \rho_{12}(t) & \rho_{13}(t) & \rho_{14}(t)\\
\rho_{21}(t) & \rho_{22}(t) & \rho_{23}(t) & \rho_{24}(t)\\
\rho_{31}(t) & \rho_{32}(t) & \rho_{33}(t) & \rho_{34}(t)\\
\rho_{41}(t) & \rho_{42}(t) & \rho_{43}(t) & \rho_{44}(t)\\
\end{pmatrix},
\label{final-rho}
\end{align}%
where $\rho_{SB}(0)$ is the joint system-bath initial state. The elements of the density matrix are given in Appendix C.

\section{Construction of the dynamical map}
Having constructed the density matrix of the reduced two-qubit central spin system, we now find the dynamical map of the system. To this effect, we derive the Kraus operators for the evolution of the reduced system. 
\subsection{Operator sum representation}
An important facet of general quantum evolution represented by a CPTP (completely positive trace preserving) operation is the Kraus operator sum representation, given as $\rho(t) = \sum_iK_i(t)\rho(0)K_i^{\dagger}(t)$. The Kraus operators, $K_i's$ can be constructed from the eigenvalues and eigenvectors of the Choi-Jamio\l{}kowski (CJ) state of the corresponding dynamical map. The CJ state of the dynamic map $\phi (\cdot)$ acting on a $d-$dimensional system is given by $({\mathbb I}_d \otimes \phi)\ketbra{\psi}{\psi}$, with $\ket{\psi}$ being the maximally entangled state in $d^2$ dimension. For our particular case, the CJ matrix is given by,

{\footnotesize
\begin{widetext}
\begin{align}
\mathcal{C}(t)=\begin{pmatrix}
	|A_1|^2&0&0&0&0&A_1J_1^*&0&0&0&0&A_1G_1^*&0&0&0&0&A_1D_1^*\\
	0&(n+1)|B_1|^2&0&0&0&0&0&0&0&0&0&0&0&0&0&0\\
	0&0&(m+1)|C_1|^2&0&0&0&0&0&0&0&0&0&0&0&0&0\\
	0&0&0&0&0&0&0&0&0&0&0&0&0&0&0&0\\
	0&0&0&0&n|K_1|^2&0&0&0&0&0&0&0&0&0&0&0\\
	A_1^*J_1&0&0&0&0&|J_1|^2&0&0&0&0&J_1G_1^*&0&0&0&0&J_1D_1^*\\
	0&0&0&0&0&0&0&0&0&0&0&0&0&0&0&0\\
	0&0&0&0&0&0&0&(m+1)|L_1|^2&0&0&0&0&0&0&0&0\\
	0&0&0&0&0&0&0&0&m|I_1|^2&0&0&0&0&0&0&0\\
	0&0&0&0&0&0&0&0&0&0&0&0&0&0&0&0\\
	A_1^*G_1&0&0&0&0&J_1^*G_1&0&0&0&0&|G_1|^2&0&0&0&0&G_1D_1^*\\
	0&0&0&0&0&0&0&0&0&0&0&(n+1)|H_1|^2&0&0&0&0\\
	0&0&0&0&0&0&0&0&0&0&0&0&0&0&0&0\\
	0&0&0&0&0&0&0&0&0&0&0&0&0&m|F_1|^2&0&0\\
	0&0&0&0&0&0&0&0&0&0&0&0&0&0&n|E_1|^2&0\\
	A_1^*D_1&0&0&0&0&J_1^*D_1&0&0&0&0&G_1^*D_1&0&0&0&0&|D_1|^2
\end{pmatrix}.
\label{choi-matrix}
\end{align}
\end{widetext}
}
The symbols and abbreviations used in Eq. (\ref{choi-matrix}) are explained in appendices A and D. One can find the eigenspectrum of the matrix in Eq. (\ref{choi-matrix}) numerically, and from there the Kraus operators for the evolution of the state can be derived. These Kraus operators should satisfy the relation $\sum_i K_i^{\dagger}K_i = \mathbb{I}$, which is indeed the case here.    

\section{Analysis of the dynamical map}
In this section, we use the dynamical map of the reduced state of the system to shed light on the dynamics of the two-qubit central spin model. Particularly we are interested in the study of its non-Markovian behaviour from the backdrop of information backflow \citep{BLP-measure}. The study of information backflow is done by observing the time evolution of certain distance functions like trace distance or fidelity between two quantum states, which is a monotonically decreasing function under Markovian evolution. Any deviation from their monotonic behavior is thus considered as a signature of non-Markovianity. Here we study the evolution of trace distance to investigate the non-Markovian nature of the dynamics. In this context, it is interesting to observe the difference between the action of the dynamical map in their local and global avatars. Along with this, we also investigate the evolution of non-classical correlations such as quantum entanglement and quantum discord, to determine the sustenance of such resourceful quantum properties under the dynamics in question.

\subsection{Trace distance as a witness of non-Markovianity}
The trace distance, which gives a measure of distinguishability between two quantum states, is given as ${\mathcal D}(\rho_1, \rho_2) = \frac{1}{2}||\rho_1 - \rho_2||_1$, where $||(\cdot)||_1 = {\rm Tr}\sqrt{(\cdot)^{\dagger}(\cdot)}$. To this end, we calculate the trace distance between the state evolved with time through the dynamical map given above and the initial state, and is given by,
\begin{align}
	{\mathcal D}(\rho_{S_1 S_2}(t), \rho_{S_1 S_2}(0)) = \frac{1}{2}||\rho_{S_1 S_2}(t) - \rho_{S_1 S_2}(0)||_1.
\end{align}
\begin{figure}[h]
	\includegraphics[width=1\columnwidth]{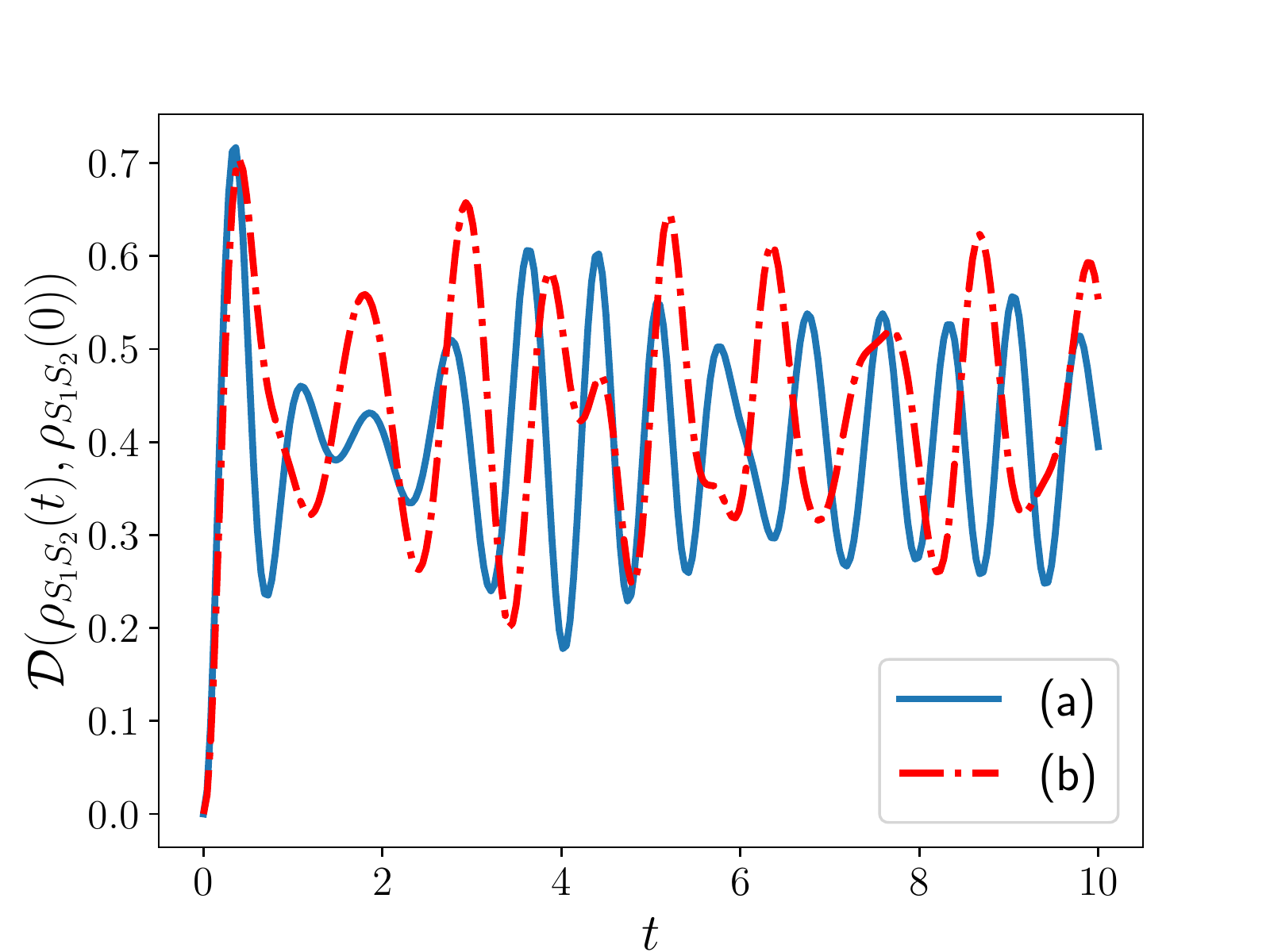}
	\caption{(Colour online) Variation of trace distance, between time evolved state and initial state, with time for the reduced state of the system. In (a), the initial state is taken to be $\ket{11}$, and in (b), the initial state is taken to be $\ket{10}$. The parameters have following values: $\omega_1 = 2.0$, $\omega_2 = 1.9$, $\delta = 2.5$, $\omega_a = 1.1$, $\omega_b = 1.2$, $M=N=100$, $T=1$, $\epsilon_1 = 2.6$, $\epsilon_2 = 2.5$.}
	\label{trace-dist-fig}
\end{figure}
In Fig. \ref{trace-dist-fig}, we can see the variation of ${\mathcal D}(\rho_{S_1 S_2}(t), \rho_{S_1 S_2}(0))$ with time for different initial states. The non monotonic behaviour of the trace distance between time evolved state and the initial state is a witness of non-Markovianity in the system.  

\subsection{Difference between local and global dynamical maps}
In this work, we have derived the global dynamical map of two central spins $(\Lambda_{12})$. Let $\Lambda_{1}$, $\Lambda_{2}$ be the local dynamical maps derived by solving the local Lindblad equations for each central spin \citep{sam2}, with the added proviso that the bath spins are interacting with each other. $\Lambda_{12}$ is the global map constructed here. We take the same parameter values for both the global and local maps, so that we can observe the difference between them from a common footing. For a particular initial state $\rho_s(0)$ we can calculate the trace distance as, 
\begin{equation}
	{\mathcal D}(\rho_{\text{global}}(t), \rho_{\text{local}}(t)) = \frac{1}{2}||\Lambda_{12}(\rho_s(0)) - \Lambda_1 \otimes \Lambda_2 (\rho_s(0))||_1. 
\end{equation}

\begin{figure}[h]
	\includegraphics[width=1\columnwidth]{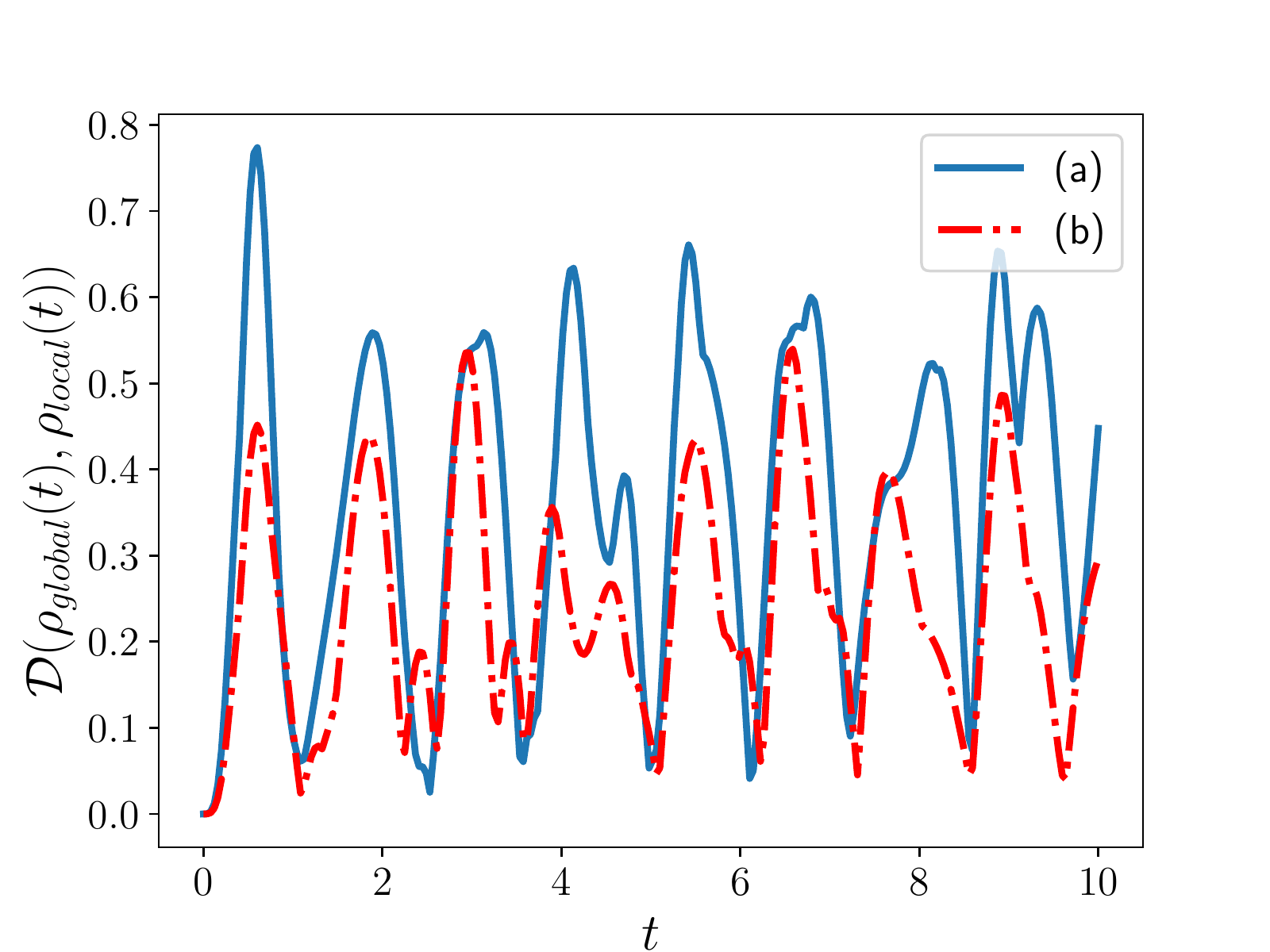}
	\caption{(Colour online) Variation of trace distance for difference between local and global maps with time for the reduced state of the system. In (a), the initial state is taken to be $\ket{11}$, and in (b), the initial state is taken to be $\ket{10}$. The parameters have following values: $\omega_1 = 2.0$, $\omega_2 = 1.9$, $\delta = 5$, $\omega_a = 1.1$, $\omega_b = 1.2$, $M=N=100$, $T=1$, $\epsilon_1 = 2.6$, $\epsilon_2 = 2.5$.}
	\label{global-v-local-fig}
\end{figure}
In Fig. \ref{global-v-local-fig}, we can see the difference between local and global dynamical maps by setting the same values for all the parameters. It is evident from the plot that though there is no interaction between the two baths in the dynamics we are evaluating, and the two baths are acting separately with individual qubits, there is a distinct difference in the dynamical behaviour of this map with the local maps acting separately with each of the qubits. Through the interaction between the qubits, bath information is passed from one environment to another and hence, despite being mutually non-interacting, the action of the bath exhibits a global trait \citep{BANERJEE-global-trait}. Therefore, it is our understanding that, in those situations where baths are acting locally on a bipartite system, as presented in Eq. \eqref{1}, applying local Kraus operators for each separate system-bath interaction, does not give us the complete picture.

\subsection{Quantum Correlations}
Quantum correlations are a very useful resource in quantum information processing. To this end, we investigate the quantum correlations in the reduced state of the system given in Eq. (\ref{final-rho}). Concurrence \citep{concurrence-wootters} is a measure of quantum entanglement in the system given by,
\begin{equation}
	\mathcal{Q_C} = \max\{0, \lambda_1 - \lambda_2 - \lambda_3 - \lambda_4\},
\end{equation}
where $\lambda_i's$ are the eigenvalue of the matrix $\sqrt{\sqrt{\rho_{S_1S_2}}\tilde{\rho}_{S_1S_2}\sqrt{\rho_{S_1S_2}}}$ in decreasing order, and $\tilde{\rho}_{S_1S_2} = (\sigma_y\otimes\sigma_y)\rho^*(\sigma_y\otimes\sigma_y)$. 
\begin{figure}[h]
	\includegraphics[width=1\columnwidth]{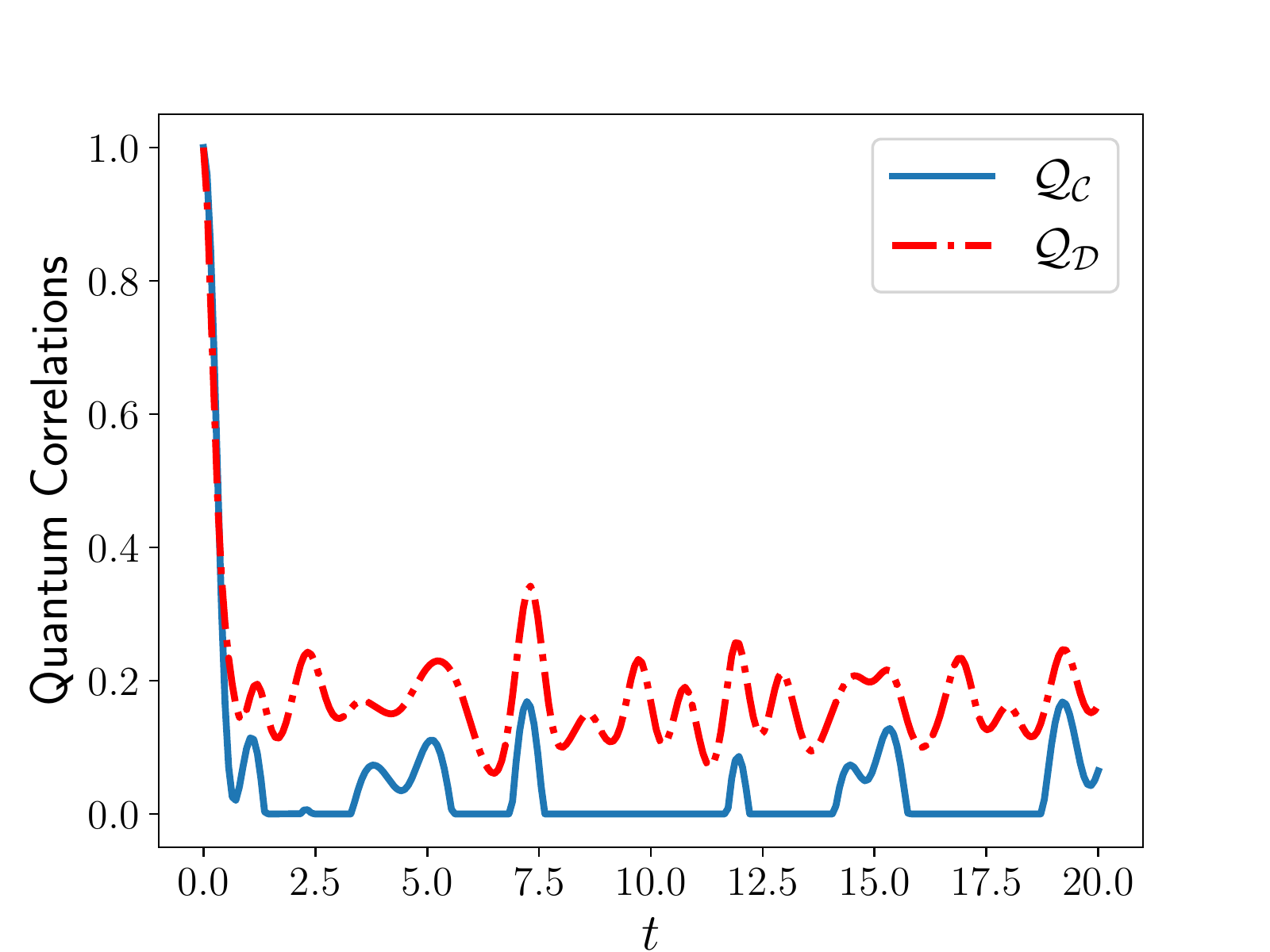}
	\caption{(Colour online) Variation of concurrence ($\mathcal{Q_C}$) and quantum discord ($\mathcal{Q_D}$) with time for the reduced state of the system. The parameters have following values: $\omega_1 = 2.0$, $\omega_2 = 1.9$, $\delta = 3$, $\omega_a = 1.1$, $\omega_b = 1.2$, $M=N=100$, $T=1$, $\epsilon_1 = 1.3$, $\epsilon_2 = 1.25$.}
	\label{quantum_correlations-fig}
\end{figure}

Another popular candidate for the measurement of quantum correlations is quantum discord \citep{quantum-discord}. It includes the quantum correlations due to quantum effects in the system which may not necessarily be due to quantum entanglement. For the reduced state given in Eq. (\ref{final-rho}), quantum discord can be given as, 
\begin{equation}
	\mathcal{Q_D} = S(\rho_{S_2}) - S(\rho_{S_1S_2}) + S(\rho_{S_1|S_2}), 
\end{equation}
where $S(\rho_{S_2})$ and $S(\rho_{S_1S_2})$ are the von Neumann entropy of the reduced subsystem $Tr_{S_1}[\rho_{S_1S_2}]$ and joint von Neumann entropy of the reduced system $\rho_{S_1S_2}$, respectively. $S(\rho_{S_1|S_2})$ is the quantum conditional entropy given by, 
\begin{equation}
	S(\rho_{S_1|S_2}) = \min_{\{\Pi_k\}}\sum_{k=1}^2 p_k S(\rho_{S_1|\Pi_k}),
\end{equation}
where $\rho_{S_1|\Pi_k}$ is the state of the reduced system when measurement operator $\Pi_k$ is operated on subsystem $S_2$, such that $\rho_{S_1|\Pi_k} = \frac{1}{p_k}Tr_{S_2}(\Pi_k\rho_{S_1S_2}\Pi_k^\dagger)$. $p_k$ is the probability associated with the measurement operators $\Pi_k$ given by $p_k = Tr(\Pi_k\rho_{S_1S_2}\Pi_k^\dagger)$. The generalized measurement operators $\Pi_k$ in 2 qubits, are given by $\Pi_1 = \mathbb{I}_{S_1}\otimes \ketbra{u}{u}_{S_2}$ and $\Pi_2 = \mathbb{I}_{S_1}\otimes \ketbra{v}{v}_{S_2}$, where $\ket{u} = \cos(\theta)\ket{1} + e^{i\phi}\sin(\theta)\ket{0}$ and $\ket{v} = \sin(\theta)\ket{1} - e^{i\phi}\cos(\theta)\ket{0}$. The parameters $\theta$ and $\phi$ vary in the range $0\le\theta\le\pi/2$ and $0\le\phi\le 2\pi$. 
We consider the two-qubit maximally entangled state, $\ket{\psi(0)}_S = \frac{1}{\sqrt{2}}(\ket{00} + \ket{11})$ as the initial state of the central spin system and study the variation of quantum correlations with time, depicted in Fig. \ref{quantum_correlations-fig}. The profile of both the quantum discord and the concurrence is similar. There exist time intervals where entanglement is zero, but discord is non-zero. Revivals in the quantum correlations depicting the information backflow nature of non-Markovianity of the system are observed. It is also important to mention that, here we are considering finite bath spins in our numerical analysis. If we extend this to the thermodynamic limit of $M,~N\rightarrow \infty$, the dynamical behavior will fall into the Markovian regime. In fact, in a previous work \citep{sam2}, one of the present authors has shown that with increasing bath spins the dynamical behavior asymptotically reaches the Markovian situation. 
Therefore, a natural conclusion is that non-Markovianity is a result of the finite-ness of the spin environment. On the other hand, to eliminate the possible conclusion that non-Markovianity is an artifact of the truncated Holstein-Primakoff transformation, we refer to a previous work on similar spin environment treated by different methods, but also exhibiting typical non-Markovian behaviour~\cite{breuer1}. This shows that non-Markovianity is not an artifact of the method.

\section{Conclusion}
In this article, we have derived the exact reduced dynamics of two coupled central spins where each spin is centrally coupled to different thermal baths. We develop the corresponding dynamical map by constructing the relevant Kraus operators. This helps to shed light into the reduced dynamics of the coupled central spin model. In particular, we show evidence of the non-Markovian evolution of the system in the form of trace distance. We also calculate the difference between the local and global dynamical maps of evolution and show that there is a distinct difference in the exact dynamics of the two qubit system interacting separately with two non-interacting environments and that of the phenomenological application of local Kraus operations of the same physical picture. It is our assertion that the later does not give us the complete picture of the exact dynamical behaviour of such interactions, where we overlook the information flow between the baths via the interaction between the qubit systems. Moreover, we have also seen the effect on the quantum correlations, in particular entanglement and quantum discord, between the central spins as the dynamics of the system evolves in time. Revival of the quantum correlations benchmarks the non-Markovian behaviour in the central spin system. 

\section*{Acknowledgements}
SB acknowledges support from Interdisciplinary Cyber Physical Systems (ICPS) programme of the Department of Science and Technology (DST),
India, Grant No.: DST/ICPS/QuST/Theme-1/2019/13. SB also acknowledges support from the Interdisciplinary Research Platform (IDRP) on Quantum Information and Computation (QIC) at IIT Jodhpur.

\appendix

\section{Derivation of the elements of the reduced state of the system.}

Following the procedure in Sec. III, we now define, $|\chi(0)\rangle= |00\rangle|xy\rangle$ and $|\chi(t)\rangle= \hat{D}(t)|00\rangle|xy\rangle + \hat{E}(t)|01\rangle|xy\rangle + \hat{F}(t)|10\rangle|xy\rangle$. Substituting $\hat{D}(t)=\hat{D_1}(t)$, $\hat{E}(t)=b \hat{E_1}(t)$ and $\hat{F}(t)=a \hat{F_1}(t)$, we find,
\begin{align}
\frac{d \hat{D_1}(t)}{dt}= &-i\Big[\frac{-\omega_1 -\omega_2 +\delta}{2} + \omega_a \Big\{\hat{m}\Big(1-\frac{\hat{m} -1}{M}\Big)-\frac{\openone}{2}\Big\} \nonumber\\
&+ \omega_b \Big\{\hat{n}\Big(1-\frac{\hat{n} -1}{N}\Big)-\frac{\openone}{2}\Big\}\Big]\hat{D_1}(t) \nonumber\\
&-i\epsilon_2\hat{n}\Big(1-\frac{\hat{n}-1}{N}\Big)^{1/2}\hat{E_1}(t) \nonumber\\
&-i\epsilon_1\hat{m}\Big(1-\frac{\hat{m}-1}{M}\Big)^{1/2}\hat{F_1}(t), \nonumber\\
\frac{d \hat{E_1}(t)}{dt}= &-i\Big[\frac{-\omega_1 +\omega_2 -\delta}{2} + \omega_a \Big\{\hat{m}\Big(1-\frac{\hat{m} -1}{M}\Big)-\frac{\openone}{2}\Big\} \nonumber\\
&+ \omega_b \Big\{(\hat{n}-1)\Big(1-\frac{\hat{n} -2}{N}\Big)-\frac{\openone}{2}\Big\}\Big]\hat{E_1}(t) \nonumber\\
&-i\epsilon_2\Big(1-\frac{\hat{n}-1}{N}\Big)^{1/2}\hat{D_1}(t), \nonumber\\
\frac{d \hat{F_1}(t)}{dt}= &-i\Big[\frac{\omega_1 -\omega_2 -\delta}{2} + \omega_a \Big\{(\hat{m}-1)\Big(1-\frac{\hat{m} -2}{M}\Big)-\frac{\openone}{2}\Big\} \nonumber\\
&+ \omega_b \Big\{\hat{n}\Big(1-\frac{\hat{n} -1}{N}\Big)-\frac{\openone}{2}\Big\}\Big]\hat{F_1}(t) \nonumber\\
&-i\epsilon_1\Big(1-\frac{\hat{m}-1}{M}\Big)^{1/2}\hat{D_1}(t).
\label{zero-zero}
\end{align}
The evolution of the reduced state ($|00\rangle\langle00|$) by tracing over bath modes can be found as,
\begin{align}
\phi(|00\rangle\langle 00|) &=\tr_{B_1 B_2}[|\chi(t)\rangle\langle\chi(t)|] \nonumber\\
&=\frac{1}{Z}\sum_m^M \sum_n^N \Big\{|D_1(m,n,t)|^2 |00\rangle\langle 00| \nonumber\\
&+ n |E_1(m,n,t)|^2 |01\rangle\langle 01| \nonumber\\
&+ m |F_1(m,n,t)|^2 |10\rangle\langle 10|\Big\} \nonumber\\
&\times \exp[-\frac{\hslash \omega_a}{K_B T}\{m\Big(1-\frac{m-1}{M}\Big)-\frac{1}{2}\}] \nonumber\\
&\times \exp[-\frac{\hslash \omega_b}{K_B T}\{n\Big(1-\frac{n-1}{N}\Big)-\frac{1}{2}\}]
\end{align}
One may find the components $D_1(m,n,t)$, $E_1(m,n,t)$ and $F_1(m,n,t)$, which are the eigenvalues of the operators, $\hat{D_1}(t)$, $\hat{E_1}(t)$ and $\hat{F_1}(t)$ with eigenvectors $|mn\rangle$, by following the steps described in the Appendix B.

In a similar manner, we specify $|\xi(0)\rangle= |01\rangle|xy\rangle$ and $|\xi(t)\rangle= \hat{G}(t)|01\rangle|xy\rangle + \hat{H}(t)|00\rangle|xy\rangle + \hat{I}(t)|11\rangle|xy\rangle$. Now substituting with, $\hat{G}(t)=\hat{G_1}(t)$, $\hat{H}(t)=b^{\dagger} \hat{H_1}(t)$ and $\hat{I}(t)=a \hat{I_1}(t)$, we can have,
\begin{align}
\frac{d \hat{G_1}(t)}{dt}= &-i\Big[\frac{-\omega_1 +\omega_2 -\delta}{2} + \omega_a \Big\{\hat{m}\Big(1-\frac{\hat{m} -1}{M}\Big)-\frac{\openone}{2}\Big\} \nonumber\\
&+ \omega_b \Big\{\hat{n}\Big(1-\frac{\hat{n} -1}{N}\Big)-\frac{\openone}{2}\Big\}\Big]\hat{G_1}(t) \nonumber\\
&-i\epsilon_2\Big(1-\frac{\hat{n}}{N}\Big)^{1/2}(\hat{n}+1)\hat{H_1}(t) \nonumber\\
&-i\epsilon_1\hat{m}\Big(1-\frac{\hat{m}-1}{M}\Big)^{1/2}\hat{I_1}(t), \nonumber\\
\frac{d \hat{H_1}(t)}{dt}= &-i\Big[\frac{-\omega_1 -\omega_2 +\delta}{2} + \omega_a \Big\{\hat{m}\Big(1-\frac{\hat{m} -1}{M}\Big)-\frac{\openone}{2}\Big\} \nonumber\\
&+ \omega_b \Big\{(\hat{n}+1)\Big(1-\frac{\hat{n}}{N}\Big)-\frac{\openone}{2}\Big\}\Big]\hat{H_1}(t) \nonumber\\
&-i\epsilon_2\Big(1-\frac{\hat{n}}{N}\Big)^{1/2}\hat{G_1}(t), \nonumber\\
\frac{d \hat{I_1}(t)}{dt}= &-i\Big[\frac{\omega_1 +\omega_2 +\delta}{2} + \omega_a \Big\{(\hat{m}-1)\Big(1-\frac{\hat{m} -2}{M}\Big)-\frac{\openone}{2}\Big\} \nonumber\\
&+ \omega_b \Big\{\hat{n}\Big(1-\frac{\hat{n} -1}{N}\Big)-\frac{\openone}{2}\Big\}\Big]\hat{I_1}(t) \nonumber\\
&-i\epsilon_1\Big(1-\frac{\hat{m}-1}{M}\Big)^{1/2}\hat{G_1}(t).
\label{zero-one}
\end{align}
We may now express the evolution of the reduced state ($|01\rangle\langle01|$) in the following way,
\begin{align}
\phi(|01\rangle\langle 01|) &=\tr_{B_1 B_2}[|\xi(t)\rangle\langle\xi(t)|] \nonumber\\
&=\frac{1}{Z}\sum_m^M \sum_n^N \Big\{|G_1(m,n,t)|^2 |01\rangle\langle 01| \nonumber\\
&+ (n+1) |H_1(m,n,t)|^2 |00\rangle\langle 00| \nonumber\\
&+ m |I_1(m,n,t)|^2 |11\rangle\langle 11|\Big\} \nonumber\\
&\times \exp[-\frac{\hslash \omega_a}{K_B T}\{m\Big(1-\frac{m-1}{M}\Big)-\frac{1}{2}\}] \nonumber\\
&\times \exp[-\frac{\hslash \omega_b}{K_B T}\{n\Big(1-\frac{n-1}{N}\Big)-\frac{1}{2}\}]
\end{align}
Following the procedure written in the Appendix B, one may calculate the eigenvalues of $\hat{G_1}(t)$, $\hat{H_1}(t)$ and $\hat{I_1}(t)$ i.e. $G_1(m,n,t)$, $H_1(m,n,t)$ and $I_1(m,n,t)$ for eigenvectors $|mn\rangle$.

To determine the evolution of the fourth diagonal element ($|10\rangle\langle10|$) of the density matrix of the system of two central spins, we assume, $|\varrho(0)\rangle= |10\rangle|xy\rangle$ and $|\varrho(t)\rangle= \hat{J}(t)|10\rangle|xy\rangle + \hat{K}(t)|11\rangle|xy\rangle + \hat{L}(t)|00\rangle|xy\rangle$. Then we replace the time-dependent operators as, $\hat{J}(t)=\hat{J_1}(t)$, $\hat{K}(t)=b \hat{K_1}(t)$ and $\hat{L}(t)=a^{\dagger} \hat{L_1}(t)$ and we obtain,
\begin{align}
\frac{d \hat{J_1}(t)}{dt}= &-i\Big[\frac{\omega_1 -\omega_2 -\delta}{2} + \omega_a \Big\{\hat{m}\Big(1-\frac{\hat{m} -1}{M}\Big)-\frac{\openone}{2}\Big\} \nonumber\\
&+ \omega_b \Big\{\hat{n}\Big(1-\frac{\hat{n} -1}{N}\Big)-\frac{\openone}{2}\Big\}\Big]\hat{J_1}(t) \nonumber\\
&-i\epsilon_2\hat{n}\Big(1-\frac{\hat{n}-1}{N}\Big)^{1/2}\hat{K_1}(t) \nonumber\\
&-i\epsilon_1\Big(1-\frac{\hat{m}}{M}\Big)^{1/2}(\hat{m}+1)\hat{L_1}(t), \nonumber\\
\frac{d \hat{K_1}(t)}{dt}= &-i\Big[\frac{\omega_1 +\omega_2 +\delta}{2} + \omega_a \Big\{\hat{m}\Big(1-\frac{\hat{m} -1}{M}\Big)-\frac{\openone}{2}\Big\} \nonumber\\
&+ \omega_b \Big\{(\hat{n}-1)\Big(1-\frac{\hat{n}-2}{N}\Big)-\frac{\openone}{2}\Big\}\Big]\hat{K_1}(t) \nonumber\\
&-i\epsilon_2\Big(1-\frac{\hat{n}-1}{N}\Big)^{1/2}\hat{J_1}(t), \nonumber\\
\frac{d \hat{L_1}(t)}{dt}= &-i\Big[\frac{-\omega_1 -\omega_2 +\delta}{2} + \omega_a \Big\{(\hat{m}+1)\Big(1-\frac{\hat{m}}{M}\Big)-\frac{\openone}{2}\Big\} \nonumber\\
&+ \omega_b \Big\{\hat{n}\Big(1-\frac{\hat{n} -1}{N}\Big)-\frac{\openone}{2}\Big\}\Big]\hat{L_1}(t) \nonumber\\
&-i\epsilon_1\Big(1-\frac{\hat{m}}{M}\Big)^{1/2}\hat{J_1}(t).
\label{one-zero}
\end{align}
The reduced state now changes as,
\begin{align}
\phi(|10\rangle\langle 10|) &=\tr_{B_1 B_2}[|\varrho(t)\rangle\langle\varrho(t)|] \nonumber\\
&=\frac{1}{Z}\sum_m^M \sum_n^N \Big\{|J_1(m,n,t)|^2 |10\rangle\langle 10| \nonumber\\
&+ n |K_1(m,n,t)|^2 |11\rangle\langle 11| \nonumber\\
&+ (m+1) |L_1(m,n,t)|^2 |00\rangle\langle 00|\Big\} \nonumber\\
&\times \exp[-\frac{\hslash \omega_a}{K_B T}\{m\Big(1-\frac{m-1}{M}\Big)-\frac{1}{2}\}] \nonumber\\
&\times \exp[-\frac{\hslash \omega_b}{K_B T}\{n\Big(1-\frac{n-1}{N}\Big)-\frac{1}{2}\}],
\end{align}
where, by definition, $\hat{J_1}(t)|mn\rangle=J_1(m,n,t)|mn\rangle$, $\hat{K_1}(t)|mn\rangle=K_1(m,n,t)|mn\rangle$ and $\hat{L_1}(t)|mn\rangle=L_1(m,n,t)|mn\rangle$. The corresponding eigenvalues can be found from Eq. (\ref{one-zero}) by following the approach mentioned in the Appendix B.

Now the off-diagonal components of the reduced density matrix will take the following forms:
\begin{align}
\phi(|11\rangle\langle 00|) &=\tr_{B_1 B_2}[|\psi(t)\rangle\langle\chi(t)|] \nonumber\\
&=\frac{1}{Z}\sum_m^M \sum_n^N \Big(A_1(m,n,t) D_1^{*}(m,n,t) |11\rangle\langle 00| \Big) \nonumber\\
&\times \exp[-\frac{\hslash \omega_a}{K_B T}\{m\Big(1-\frac{m-1}{M}\Big)-\frac{1}{2}\}] \nonumber\\
&\times \exp[-\frac{\hslash \omega_b}{K_B T}\{n\Big(1-\frac{n-1}{N}\Big)-\frac{1}{2}\}],
\end{align}
\begin{align}
\phi(|11\rangle\langle 01|) &=\tr_{B_1 B_2}[|\psi(t)\rangle\langle\xi(t)|] \nonumber\\
&=\frac{1}{Z}\sum_m^M \sum_n^N \Big(A_1(m,n,t) G_1^{*}(m,n,t) |11\rangle\langle 01| \Big) \nonumber\\
&\times \exp[-\frac{\hslash \omega_a}{K_B T}\{m\Big(1-\frac{m-1}{M}\Big)-\frac{1}{2}\}] \nonumber\\
\bibliographystyle{}&\times \exp[-\frac{\hslash \omega_b}{K_B T}\{n\Big(1-\frac{n-1}{N}\Big)-\frac{1}{2}\}],
\end{align}
\begin{align}
\phi(|11\rangle\langle 10|) &=\tr_{B_1 B_2}[|\psi(t)\rangle\langle\varrho(t)|] \nonumber\\
&=\frac{1}{Z}\sum_m^M \sum_n^N \Big(A_1(m,n,t) J_1^{*}(m,n,t) |11\rangle\langle 10| \Big) \nonumber\\
&\times \exp[-\frac{\hslash \omega_a}{K_B T}\{m\Big(1-\frac{m-1}{M}\Big)-\frac{1}{2}\}] \nonumber\\
&\times \exp[-\frac{\hslash \omega_b}{K_B T}\{n\Big(1-\frac{n-1}{N}\Big)-\frac{1}{2}\}],
\end{align}
\begin{align}
\phi(|10\rangle\langle 00|) &=\tr_{B_1 B_2}[|\varrho(t)\rangle\langle\chi(t)|] \nonumber\\
&=\frac{1}{Z}\sum_m^M \sum_n^N \Big(J_1(m,n,t) D_1^{*}(m,n,t) |10\rangle\langle 00| \Big) \nonumber\\
&\times \exp[-\frac{\hslash \omega_a}{K_B T}\{m\Big(1-\frac{m-1}{M}\Big)-\frac{1}{2}\}] \nonumber\\
&\times \exp[-\frac{\hslash \omega_b}{K_B T}\{n\Big(1-\frac{n-1}{N}\Big)-\frac{1}{2}\}],
\end{align}
\begin{align}
\phi(|10\rangle\langle 01|) &=\tr_{B_1 B_2}[|\varrho(t)\rangle\langle\xi(t)|] \nonumber\\
&=\frac{1}{Z}\sum_m^M \sum_n^N \Big(J_1(m,n,t) G_1^{*}(m,n,t) |10\rangle\langle 01| \Big) \nonumber\\
&\times \exp[-\frac{\hslash \omega_a}{K_B T}\{m\Big(1-\frac{m-1}{M}\Big)-\frac{1}{2}\}] \nonumber\\
&\times \exp[-\frac{\hslash \omega_b}{K_B T}\{n\Big(1-\frac{n-1}{N}\Big)-\frac{1}{2}\}],
\end{align}
\begin{align}
\phi(|01\rangle\langle 00|) &=\tr_{B_1 B_2}[|\xi(t)\rangle\langle\chi(t)|] \nonumber\\
&=\frac{1}{Z}\sum_m^M \sum_n^N \Big(G_1(m,n,t) D_1^{*}(m,n,t) |01\rangle\langle 00| \Big) \nonumber\\
&\times \exp[-\frac{\hslash \omega_a}{K_B T}\{m\Big(1-\frac{m-1}{M}\Big)-\frac{1}{2}\}] \nonumber\\
&\times \exp[-\frac{\hslash \omega_b}{K_B T}\{n\Big(1-\frac{n-1}{N}\Big)-\frac{1}{2}\}].
\end{align}

From the hermiticity of the reduced density matrix, the other off-diagonal elements can be expressed as,
\begin{align*}
\phi(|00\rangle\langle 11|) = \Big(\phi(|11\rangle\langle 00|)\Big)^{\dagger},
\phi(|01\rangle\langle 11|) = \Big(\phi(|11\rangle\langle 01|)\Big)^{\dagger}, \\
\phi(|10\rangle\langle 11|) = \Big(\phi(|11\rangle\langle 10|)\Big)^{\dagger},
\phi(|00\rangle\langle 10|) = \Big(\phi(|10\rangle\langle 00|)\Big)^{\dagger}, \\
\phi(|01\rangle\langle 10|) = \Big(\phi(|10\rangle\langle 01|)\Big)^{\dagger},
\phi(|00\rangle\langle 01|) = \Big(\phi(|01\rangle\langle 00|)\Big)^{\dagger}.
\end{align*}

\section{Solution of simultaneous linear differential equations from boundary conditions: Matrix method}

Suppose, three simultaneous linear differential equations are written as:
\begin{align}
&\frac{d X(t)}{dt}= -i a X(t) -i d Y(t) -i e Z(t), \nonumber\\
&\frac{d Y(t)}{dt}= -i f X(t) -i b Y(t), \nonumber\\
&\frac{d Z(t)}{dt}= -i g X(t) -i c Z(t).
\label{Diff}
\end{align}
And we have to find the solution using the given boundary conditions: $X(0)=1$, $Y(0)=0$ and $Z(0)=0$.

The problem can be solved by plugging in the well-known Matrix method in this context. According to this, one can write Eq. (\ref{Diff}) using column and square matrices of dimension 3 in a following way,
\begin{eqnarray}
\frac{d}{dt}\begin{pmatrix}
X(t)\\
Y(t)\\
Z(t)\\
\end{pmatrix}= -i \begin{pmatrix}
a & d & e\\
f & b & 0\\
g & 0 & c\\
\end{pmatrix} \begin{pmatrix}
X(t)\\
Y(t)\\
Z(t)\\
\end{pmatrix}
\end{eqnarray}
For simplicity, we denote, $M$=$\begin{pmatrix}
a & d & e\\
f & b & 0\\
g & 0 & c\\
\end{pmatrix}$. Any of the three eigenvalues, say $\lambda_j$($j$=1,2,3) of matrix, $M$ can be found by solving the characteristic Eq,
\begin{align}
(a-\lambda_j) (b-\lambda_j) (c-\lambda_j) -f d (c-\lambda_j) - e g (b-\lambda_j)=0, ~ \forall j.
\label{characteristic}
\end{align}
And corresponding to each $\lambda_j$, eigenvectors can be expressed as, $\begin{pmatrix}
(\lambda_j -b) (\lambda_j -c)\\
f (\lambda_j -c)\\
g (\lambda_j -b)\\
\end{pmatrix}\equiv \begin{pmatrix}
\alpha_j\\
\beta_j\\
\gamma_j\\
\end{pmatrix}$, where, $\alpha_j=(\lambda_j -b) (\lambda_j -c)$, $\beta_j=f (\lambda_j -c)$ and $\gamma_j=g (\lambda_j -b)$, except for the case when $\lambda_j=b,\;c$.
Then according to the Matrix method, the general solution of Eq. (\ref{Diff}) in terms of column matrix will take the form,
\begin{eqnarray}
\begin{pmatrix}
X(t)\\
Y(t)\\
Z(t)\\
\end{pmatrix}= \sum_{j=1}^3 v_j ~ e^{-i \lambda_j t} ~ \begin{pmatrix}
\alpha_j\\
\beta_j\\
\gamma_j\\
\end{pmatrix},
\label{soln}
\end{eqnarray}
where the co-efficients $v_j$ can be computed from three boundary conditions given as,
\begin{eqnarray}
\sum_{j=1}^3 v_j \alpha_j =1;~~ \sum_{j=1}^3 v_j \beta_j =0;~~ \sum_{j=1}^3 v_j \gamma_j =0.
\label{boundary}
\end{eqnarray}
Now one can write from Eq. (\ref{boundary}),
\begin{align}
&v_1 = \frac{\beta_3 \gamma_2 - \beta_2 \gamma_3}{\alpha_1 (\beta_3 \gamma_2 - \beta_2 \gamma_3) + \alpha_2 (\beta_1 \gamma_3 - \beta_3 \gamma_1) + \alpha_3 (\beta_2 \gamma_1 - \beta_1 \gamma_2)}, \nonumber\\
&v_2 = \frac{\beta_1 \gamma_3 - \beta_3 \gamma_1}{\alpha_1 (\beta_3 \gamma_2 - \beta_2 \gamma_3) + \alpha_2 (\beta_1 \gamma_3 - \beta_3 \gamma_1) + \alpha_3 (\beta_2 \gamma_1 - \beta_1 \gamma_2)}, \nonumber\\
&v_3 = \frac{\beta_2 \gamma_1 - \beta_1 \gamma_2}{\alpha_1 (\beta_3 \gamma_2 - \beta_2 \gamma_3) + \alpha_2 (\beta_1 \gamma_3 - \beta_3 \gamma_1) + \alpha_3 (\beta_2 \gamma_1 - \beta_1 \gamma_2)}.
\end{align}
We see that, we are able to come up with the solution (\ref{soln}) only when $\lambda_j$ is calculated from the Eq. (\ref{characteristic}). For solving the three roots of such cubic equation, we follow the procedure given below.

Let us say, a cubic equation is in the form:
\begin{eqnarray}
a_1 x^3 + b_1 x^2 + c_1 x + d_1 =0
\label{cubic}
\end{eqnarray}
Then we define the following quantities,
\begin{align*}
&\Delta= 18 a_1 b_1 c_1 d_1 - 4 b_1^3 d_1 + b_1^2 c_1^2 - 4 a_1 c_1^3 - 27 a_1^2 d_1^2, \nonumber\\
&\Delta_0= b_1^2 - 3 a_1 c_1, \nonumber\\
&\Delta_1 = 2 b_1^3 - 9 a_1 b_1 c_1 + 27 a_1^2 d_1, \nonumber\\
&Q= \sqrt[3]{\frac{\Delta_1 + \sqrt{\Delta_1^2 - 4\Delta_0^3}}{2}} = \sqrt[3]{\frac{\Delta_1 + \sqrt{-27 a_1^2 \Delta}}{2}}.
\end{align*}
where, $\Delta$ is called the discriminant of the cubic Eq. (\ref{cubic}). Next, denoting a complex number, $-\frac{1}{2}+\frac{\sqrt{3}}{2}i =\varphi$, we may write the roots of Eq. (\ref{cubic}) as,
\begin{eqnarray}
x_k = -\frac{1}{3 a_1}(b_1 + \varphi^k Q + \frac{\Delta_0}{\varphi^k Q}), ~ k\in \{1,2,3\}.
\end{eqnarray}
Note that, the root $x_1$ is always real and the roots $x_2$ and $x_3$ are complex and conjugate to each other only when $\Delta<0$. Otherwise, all the roots are real and they become equal when $\Delta=0$.

In our case, expanding the Left Hand Side of Eq. (\ref{characteristic}) in a power series of $\lambda_j$ for all $j$, we identify, $a_1=1$, $b_1=-(a+b+c)$, $c_1= a b + b c + a c - f d - e g$ and $d_1= f d c + e g b - a b c$. As described above, $\lambda_j$'s($j$=1,2,3) can be determined and the solution (\ref{soln}) can be attained accordingly.

All the Eqs. (\ref{one-one},\ref{zero-zero},\ref{zero-one},\ref{one-zero}) are in the form of Eq. (\ref{Diff}). Hence, all the co-efficients can be obtained in terms of $\omega_1$, $\omega_2$, $\delta$, $\omega_a$, $\omega_b$, $\epsilon_1$, $\epsilon_2$, m, n, M and N. 

Explicitly, the co-efficients, $A_1(m,n,t)$, $B_1(m,n,t)$ and $C_1(m,n,t)$ can be obtained by solving Eq. (\ref{one-one}) which in comparison to Eq. (\ref{Diff}) shows that,
\begin{align*}
X(t)&=A_1(m,n,t), \nonumber\\
Y(t)&=B_1(m,n,t), \nonumber\\
Z(t)&=C_1(m,n,t), \nonumber\\
a=&\frac{\omega_1 +\omega_2 +\delta}{2} + \omega_a \Big\{m\Big(1-\frac{m -1}{M}\Big)-\frac{1}{2}\Big\}\\
&+ \omega_b \Big\{n\Big(1-\frac{n -1}{N}\Big)-\frac{1}{2}\Big\}, \nonumber\\
b=&\frac{\omega_1 -\omega_2 -\delta}{2} + \omega_a \Big\{m\Big(1-\frac{m-1}{M}\Big)-\frac{1}{2}\Big\}\\
&+ \omega_b \Big\{(n+1)\Big(1-\frac{n-1}{N}\Big)-\frac{1}{2}\Big\}, \nonumber\\
c=&\frac{-\omega_1 +\omega_2 -\delta}{2} + \omega_a \Big\{(m+1)\Big(1-\frac{m-1}{M}\Big)-\frac{1}{2}\Big\}\\
&+ \omega_b \Big\{n\Big(1-\frac{n-1}{N}\Big)-\frac{1}{2}\Big\}, \nonumber\\
d=&\epsilon_2\Big(1-\frac{n}{N}\Big)^{1/2}(n+1), \nonumber\\
e=&\epsilon_1\Big(1-\frac{m}{M}\Big)^{1/2}(m+1), \nonumber\\
f=&\epsilon_2\Big(1-\frac{n}{N}\Big)^{1/2}, \nonumber\\
g=&\epsilon_1\Big(1-\frac{m}{M}\Big)^{1/2}.
\end{align*}
Likewise, comparing Eq. (\ref{zero-zero}) with Eq. (\ref{Diff}), we get,
\begin{align*}
X(t)&=D_1(m,n,t), \nonumber\\
Y(t)&=E_1(m,n,t), \nonumber\\
Z(t)&=F_1(m,n,t), \nonumber\\
a=&\frac{-\omega_1 -\omega_2 +\delta}{2} + \omega_a \Big\{m\Big(1-\frac{m-1}{M}\Big)-\frac{1}{2}\Big\}\\
&+ \omega_b \Big\{n\Big(1-\frac{n-1}{N}\Big)-\frac{1}{2}\Big\}, \nonumber\\
b=&\frac{-\omega_1 +\omega_2 -\delta}{2} + \omega_a \Big\{m\Big(1-\frac{m-1}{M}\Big)-\frac{1}{2}\Big\}\\
&+ \omega_b \Big\{(n-1)\Big(1-\frac{n-2}{N}\Big)-\frac{1}{2}\Big\}, \nonumber\\
c=&\frac{\omega_1 -\omega_2 -\delta}{2} + \omega_a \Big\{(m-1)\Big(1-\frac{m-2}{M}\Big)-\frac{1}{2}\Big\}\\
&+ \omega_b \Big\{n\Big(1-\frac{n-1}{N}\Big)-\frac{1}{2}\Big\}, \nonumber\\
d=&\epsilon_2n\Big(1-\frac{n-1}{N}\Big)^{1/2}, \nonumber\\
e=&\epsilon_1m\Big(1-\frac{m-1}{M}\Big)^{1/2}, \nonumber\\
f=&\epsilon_2\Big(1-\frac{n-1}{N}\Big)^{1/2}, \nonumber\\
g=&\epsilon_1\Big(1-\frac{m-1}{M}\Big)^{1/2}.
\end{align*}
Correspondingly, Eq. (\ref{zero-one}) exhibits,
\begin{align*}
X(t)&=G_1(m,n,t), \nonumber\\
Y(t)&=H_1(m,n,t), \nonumber\\
Z(t)&=I_1(m,n,t), \nonumber\\
a=&\frac{-\omega_1 +\omega_2 -\delta}{2} + \omega_a \Big\{m\Big(1-\frac{m-1}{M}\Big)-\frac{1}{2}\Big\}\\
&+ \omega_b \Big\{n\Big(1-\frac{n-1}{N}\Big)-\frac{1}{2}\Big\}, \nonumber\\
b=&\frac{-\omega_1 -\omega_2 +\delta}{2} + \omega_a \Big\{m\Big(1-\frac{m-1}{M}\Big)-\frac{1}{2}\Big\}\\
&+ \omega_b \Big\{(n+1)\Big(1-\frac{n}{N}\Big)-\frac{1}{2}\Big\}, \nonumber\\
c=&\frac{\omega_1 +\omega_2 +\delta}{2} + \omega_a \Big\{(m-1)\Big(1-\frac{m-2}{M}\Big)-\frac{1}{2}\Big\}\\
&+ \omega_b \Big\{n\Big(1-\frac{n-1}{N}\Big)-\frac{1}{2}\Big\}, \nonumber\\
d=&\epsilon_2\Big(1-\frac{n}{N}\Big)^{1/2}(n+1), \nonumber\\
e=&\epsilon_1m\Big(1-\frac{m-1}{M}\Big)^{1/2}, \nonumber\\
f=&\epsilon_2\Big(1-\frac{n}{N}\Big)^{1/2}, \nonumber\\
g=&\epsilon_1\Big(1-\frac{m-1}{M}\Big)^{1/2}.
\end{align*}
And finally the resemblance between Eq. (\ref{one-zero}) and Eq. (\ref{Diff}) manifests,
\begin{align*}
X(t)&=J_1(m,n,t), \nonumber\\
Y(t)&=K_1(m,n,t), \nonumber\\
Z(t)& =L_1(m,n,t), \nonumber\\
a=&\frac{\omega_1 -\omega_2 -\delta}{2} + \omega_a \Big\{m\Big(1-\frac{m-1}{M}\Big)-\frac{1}{2}\Big\}\\
&+ \omega_b \Big\{n\Big(1-\frac{n-1}{N}\Big)-\frac{1}{2}\Big\}, \nonumber\\
b=&\frac{\omega_1 +\omega_2 +\delta}{2} + \omega_a \Big\{m\Big(1-\frac{m-1}{M}\Big)-\frac{1}{2}\Big\}\\
&+ \omega_b \Big\{(n-1)\Big(1-\frac{n-2}{N}\Big)-\frac{1}{2}\Big\}, \nonumber\\
c=&\frac{-\omega_1 -\omega_2 +\delta}{2} + \omega_a \Big\{(m+1)\Big(1-\frac{m}{M}\Big)-\frac{1}{2}\Big\}\\
&+ \omega_b \Big\{n\Big(1-\frac{n-1}{N}\Big)-\frac{1}{2}\Big\}, \nonumber\\
d=&\epsilon_2n\Big(1-\frac{n-1}{N}\Big)^{1/2}, \nonumber\\
e=&\epsilon_1\Big(1-\frac{m}{M}\Big)^{1/2}(m+1), \nonumber\\
f=&\epsilon_2\Big(1-\frac{n-1}{N}\Big)^{1/2}, \nonumber\\
g=&\epsilon_1\Big(1-\frac{m}{M}\Big)^{1/2}.
\end{align*}
In this way, all the co-efficients can be obtained in order to reveal the complete dynamical map of the two-spin system.

\section{Matrix elements of the reduced state of the system}

The elements of the density matrix given in Eq. (\ref{final-rho}) are given by, 
\begin{align}
	\rho_{11} &= \frac{1}{Z}\sum_m^M\sum_n^N \Big(|A_1(m,n,t)|^2\rho_{11}(0) + n|K_1(m,n,t)|^2\rho_{22}(0)\nonumber\\
	&+ m|I_1(m,n,t)|^2\rho_{33}(0)\Big)\nonumber\\
	&\times \exp[-\frac{\hslash \omega_a}{K_B T}\{m\Big(1-\frac{m-1}{M}\Big)-\frac{1}{2}\}] \nonumber\\
	&\times \exp[-\frac{\hslash \omega_b}{K_B T}\{n\Big(1-\frac{n-1}{N}\Big)-\frac{1}{2}\}],
\end{align}	
\begin{align} 
	\rho_{22} &= \frac{1}{Z}\sum_m^M\sum_n^N[(n+1)\Big(B_1(m,n,t)|^2\rho_{11}(0) + |J_1(m,n,t)|^2\rho_{22}(0) \nonumber \\ 
	&+ m|F_1(m,n,t)|^2\rho_{44}(0)\Big)\nonumber\\
	&\times \exp[-\frac{\hslash \omega_a}{K_B T}\{m\Big(1-\frac{m-1}{M}\Big)-\frac{1}{2}\}] \nonumber\\
	&\times \exp[-\frac{\hslash \omega_b}{K_B T}\{n\Big(1-\frac{n-1}{N}\Big)-\frac{1}{2}\}],
\end{align}
\begin{align}
	\rho_{33} &= \frac{1}{Z}\sum_m^M\sum_n^N \Big((m+1)|C_1(m,n,t)|^2\rho_{11}(0) + |G_1(m,n,t)|^2\rho_{33}(0) \nonumber\\
	&+ n|E_1(m,n,t)|^2\rho_{44}(0)\Big)\nonumber\\
	&\times \exp[-\frac{\hslash \omega_a}{K_B T}\{m\Big(1-\frac{m-1}{M}\Big)-\frac{1}{2}\}] \nonumber\\
	&\times \exp[-\frac{\hslash \omega_b}{K_B T}\{n\Big(1-\frac{n-1}{N}\Big)-\frac{1}{2}\}],
\end{align}
\begin{align}
	\rho_{44} &= \frac{1}{Z}\sum_m^M\sum_n^N \Big((m+1)|B_1(m,n,t)|^2\rho_{22}(0) + |D_1(m,n,t)|^2\rho_{44}(0) \nonumber\\ 
	&+ (n+1)|H_1(m,n,t)|^2\rho_{33}(0)\Big)\nonumber\\
	&\times \exp[-\frac{\hslash \omega_a}{K_B T}\{m\Big(1-\frac{m-1}{M}\Big)-\frac{1}{2}\}] \nonumber\\
	&\times \exp[-\frac{\hslash \omega_b}{K_B T}\{n\Big(1-\frac{n-1}{N}\Big)-\frac{1}{2}\}],
\end{align}
\begin{align}
	\rho_{12} &=\frac{1}{Z}\sum_m^M \sum_n^N \Big(A_1(m,n,t) J_1^{*}(m,n,t) \rho_{12}(0) \Big) \nonumber\\
	&\times \exp[-\frac{\hslash \omega_a}{K_B T}\{m\Big(1-\frac{m-1}{M}\Big)-\frac{1}{2}\}] \nonumber\\
	&\times \exp[-\frac{\hslash \omega_b}{K_B T}\{n\Big(1-\frac{n-1}{N}\Big)-\frac{1}{2}\}],
\end{align}
\begin{align}
	\rho_{13} &=\frac{1}{Z}\sum_m^M \sum_n^N \Big(A_1(m,n,t) G_1^{*}(m,n,t) \rho_{13}(0) \Big) \nonumber\\
	&\times \exp[-\frac{\hslash \omega_a}{K_B T}\{m\Big(1-\frac{m-1}{M}\Big)-\frac{1}{2}\}] \nonumber\\
	&\times \exp[-\frac{\hslash \omega_b}{K_B T}\{n\Big(1-\frac{n-1}{N}\Big)-\frac{1}{2}\}],
\end{align}
\begin{align}
	\rho_{14} &=\frac{1}{Z}\sum_m^M \sum_n^N \Big(A_1(m,n,t) D_1^{*}(m,n,t) \rho_{14}(0) \Big) \nonumber\\
	&\times \exp[-\frac{\hslash \omega_a}{K_B T}\{m\Big(1-\frac{m-1}{M}\Big)-\frac{1}{2}\}] \nonumber\\
	&\times \exp[-\frac{\hslash \omega_b}{K_B T}\{n\Big(1-\frac{n-1}{N}\Big)-\frac{1}{2}\}],
\end{align}
\begin{align}
	\rho_{23} &=\frac{1}{Z}\sum_m^M \sum_n^N \Big(J_1(m,n,t) G_1^{*}(m,n,t) \rho_{23}(0) \Big) \nonumber\\
	&\times \exp[-\frac{\hslash \omega_a}{K_B T}\{m\Big(1-\frac{m-1}{M}\Big)-\frac{1}{2}\}] \nonumber\\
	&\times \exp[-\frac{\hslash \omega_b}{K_B T}\{n\Big(1-\frac{n-1}{N}\Big)-\frac{1}{2}\}],
\end{align}
\begin{align}
	\rho_{24} &=\frac{1}{Z}\sum_m^M \sum_n^N \Big(J_1(m,n,t) D_1^{*}(m,n,t) \rho_{24}(0) \Big) \nonumber\\
	&\times \exp[-\frac{\hslash \omega_a}{K_B T}\{m\Big(1-\frac{m-1}{M}\Big)-\frac{1}{2}\}] \nonumber\\
	&\times \exp[-\frac{\hslash \omega_b}{K_B T}\{n\Big(1-\frac{n-1}{N}\Big)-\frac{1}{2}\}],
\end{align}
\begin{align}
	\rho_{34} &=\frac{1}{Z}\sum_m^M \sum_n^N \Big(G_1(m,n,t) D_1^{*}(m,n,t) \rho_{34}(0) \Big) \nonumber\\
	&\times \exp[-\frac{\hslash \omega_a}{K_B T}\{m\Big(1-\frac{m-1}{M}\Big)-\frac{1}{2}\}] \nonumber\\
	&\times \exp[-\frac{\hslash \omega_b}{K_B T}\{n\Big(1-\frac{n-1}{N}\Big)-\frac{1}{2}\}].
\end{align}

\section{Elements of the Choi-Jamio\l{}kowski matrix}

We have used following abbreviations in the CJ matrix given in Eq. (\ref{choi-matrix}),
\begin{align}
	|A_1|^2 &= \frac{1}{Z}\sum_{m}^M\sum_n^N |A_1(m,n,t)|^2 \nonumber\\
	&\times \exp[-\frac{\hslash \omega_a}{K_B T}\{m\Big(1-\frac{m-1}{M}\Big)-\frac{1}{2}\}] \nonumber\\
	&\times \exp[-\frac{\hslash \omega_b}{K_B T}\{n\Big(1-\frac{n-1}{N}\Big)-\frac{1}{2}\}],
\end{align}	
\begin{align}
	(n+1)|B_1|^2 &=\frac{1}{Z}\sum_{m}^M\sum_n^N (n+1) |B_1(m,n,t)|^2 \nonumber\\
	&\times \exp[-\frac{\hslash \omega_a}{K_B T}\{m\Big(1-\frac{m-1}{M}\Big)-\frac{1}{2}\}] \nonumber\\
	&\times \exp[-\frac{\hslash \omega_b}{K_B T}\{n\Big(1-\frac{n-1}{N}\Big)-\frac{1}{2}\}],
\end{align}	
\begin{align}
	(m+1)|C_1|^2&= \frac{1}{Z}\sum_{m}^M\sum_n^N (m+1) |C_1(m,n,t)|^2\nonumber\\
	&\times \exp[-\frac{\hslash \omega_a}{K_B T}\{m\Big(1-\frac{m-1}{M}\Big)-\frac{1}{2}\}] \nonumber\\
	&\times \exp[-\frac{\hslash \omega_b}{K_B T}\{n\Big(1-\frac{n-1}{N}\Big)-\frac{1}{2}\}],
\end{align}	
\begin{align}
	n|K_1|^2&= \frac{1}{Z}\sum_{m}^M\sum_n^N n |K_1(m,n,t)|^2\nonumber\\
	&\times \exp[-\frac{\hslash \omega_a}{K_B T}\{m\Big(1-\frac{m-1}{M}\Big)-\frac{1}{2}\}] \nonumber\\
	&\times \exp[-\frac{\hslash \omega_b}{K_B T}\{n\Big(1-\frac{n-1}{N}\Big)-\frac{1}{2}\}],
\end{align}	
\begin{align}
	|J_1|^2 &= \frac{1}{Z}\sum_{m}^M\sum_n^N |J_1(m,n,t)|^2 \nonumber\\
	&\times \exp[-\frac{\hslash \omega_a}{K_B T}\{m\Big(1-\frac{m-1}{M}\Big)-\frac{1}{2}\}] \nonumber\\
	&\times \exp[-\frac{\hslash \omega_b}{K_B T}\{n\Big(1-\frac{n-1}{N}\Big)-\frac{1}{2}\}],
\end{align}	
\begin{align}
	(m+1)|L_1|^2&= \frac{1}{Z}\sum_{m}^M\sum_n^N (m+1) |L_1(m,n,t)|^2\nonumber\\
	&\times \exp[-\frac{\hslash \omega_a}{K_B T}\{m\Big(1-\frac{m-1}{M}\Big)-\frac{1}{2}\}] \nonumber\\
	&\times \exp[-\frac{\hslash \omega_b}{K_B T}\{n\Big(1-\frac{n-1}{N}\Big)-\frac{1}{2}\}],
\end{align}	
\begin{align}
	m|I_1|^2&= \frac{1}{Z}\sum_{m}^M\sum_n^N m |I_1(m,n,t)|^2\nonumber\\
	&\times \exp[-\frac{\hslash \omega_a}{K_B T}\{m\Big(1-\frac{m-1}{M}\Big)-\frac{1}{2}\}] \nonumber\\
	&\times \exp[-\frac{\hslash \omega_b}{K_B T}\{n\Big(1-\frac{n-1}{N}\Big)-\frac{1}{2}\}],
\end{align}
\begin{align}
	|G_1|^2 &= \frac{1}{Z}\sum_{m}^M\sum_n^N |G_1(m,n,t)|^2 \nonumber\\
	&\times \exp[-\frac{\hslash \omega_a}{K_B T}\{m\Big(1-\frac{m-1}{M}\Big)-\frac{1}{2}\}] \nonumber\\
	&\times \exp[-\frac{\hslash \omega_b}{K_B T}\{n\Big(1-\frac{n-1}{N}\Big)-\frac{1}{2}\}],
\end{align}
\begin{align}
	(n+1)|H_1|^2 &= \frac{1}{Z}\sum_{m}^M\sum_n^N (n+1)|H_1(m,n,t)|^2 \nonumber\\
	&\times \exp[-\frac{\hslash \omega_a}{K_B T}\{m\Big(1-\frac{m-1}{M}\Big)-\frac{1}{2}\}] \nonumber\\
	&\times \exp[-\frac{\hslash \omega_b}{K_B T}\{n\Big(1-\frac{n-1}{N}\Big)-\frac{1}{2}\}],
\end{align}
\begin{align}
	m|F_1|^2 &= \frac{1}{Z}\sum_{m}^M\sum_n^N m|F_1(m,n,t)|^2 \nonumber\\
	&\times \exp[-\frac{\hslash \omega_a}{K_B T}\{m\Big(1-\frac{m-1}{M}\Big)-\frac{1}{2}\}] \nonumber\\
	&\times \exp[-\frac{\hslash \omega_b}{K_B T}\{n\Big(1-\frac{n-1}{N}\Big)-\frac{1}{2}\}],
\end{align}
\begin{align}
	n|E_1|^2 &= \frac{1}{Z}\sum_{m}^M\sum_n^N n|E_1(m,n,t)|^2 \nonumber\\
	&\times \exp[-\frac{\hslash \omega_a}{K_B T}\{m\Big(1-\frac{m-1}{M}\Big)-\frac{1}{2}\}] \nonumber\\
	&\times \exp[-\frac{\hslash \omega_b}{K_B T}\{n\Big(1-\frac{n-1}{N}\Big)-\frac{1}{2}\}],
\end{align}
\begin{align}
	|D_1|^2 &= \frac{1}{Z}\sum_{m}^M\sum_n^N |D_1(m,n,t)|^2 \nonumber\\
	&\times \exp[-\frac{\hslash \omega_a}{K_B T}\{m\Big(1-\frac{m-1}{M}\Big)-\frac{1}{2}\}] \nonumber\\
	&\times \exp[-\frac{\hslash \omega_b}{K_B T}\{n\Big(1-\frac{n-1}{N}\Big)-\frac{1}{2}\}],
\end{align}
\begin{align}
	A_1J_1^* &= \frac{1}{Z}\sum_m^M\sum_n^N A_1(m,n,t)J_1^*(m,n,t)\nonumber\\
	&\times \exp[-\frac{\hslash \omega_a}{K_B T}\{m\Big(1-\frac{m-1}{M}\Big)-\frac{1}{2}\}] \nonumber\\
	&\times \exp[-\frac{\hslash \omega_b}{K_B T}\{n\Big(1-\frac{n-1}{N}\Big)-\frac{1}{2}\}],
\end{align}
\begin{align}
	A_1G_1^* &= \frac{1}{Z}\sum_m^M\sum_n^N A_1(m,n,t)G_1^*(m,n,t)\nonumber\\
	&\times \exp[-\frac{\hslash \omega_a}{K_B T}\{m\Big(1-\frac{m-1}{M}\Big)-\frac{1}{2}\}] \nonumber\\
	&\times \exp[-\frac{\hslash \omega_b}{K_B T}\{n\Big(1-\frac{n-1}{N}\Big)-\frac{1}{2}\}],
\end{align}
\begin{align}
	A_1D_1^* &= \frac{1}{Z}\sum_m^M\sum_n^N A_1(m,n,t)D_1^*(m,n,t)\nonumber\\
	&\times \exp[-\frac{\hslash \omega_a}{K_B T}\{m\Big(1-\frac{m-1}{M}\Big)-\frac{1}{2}\}] \nonumber\\
	&\times \exp[-\frac{\hslash \omega_b}{K_B T}\{n\Big(1-\frac{n-1}{N}\Big)-\frac{1}{2}\}],
\end{align}
\begin{align}
	J_1G_1^* &= \frac{1}{Z}\sum_m^M\sum_n^N J_1(m,n,t)G_1^*(m,n,t)\nonumber\\
	&\times \exp[-\frac{\hslash \omega_a}{K_B T}\{m\Big(1-\frac{m-1}{M}\Big)-\frac{1}{2}\}] \nonumber\\
	&\times \exp[-\frac{\hslash \omega_b}{K_B T}\{n\Big(1-\frac{n-1}{N}\Big)-\frac{1}{2}\}],
\end{align}
\begin{align}
	J_1D_1^* &= \frac{1}{Z}\sum_m^M\sum_n^N J_1(m,n,t)D_1^*(m,n,t)\nonumber\\
	&\times \exp[-\frac{\hslash \omega_a}{K_B T}\{m\Big(1-\frac{m-1}{M}\Big)-\frac{1}{2}\}] \nonumber\\
	&\times \exp[-\frac{\hslash \omega_b}{K_B T}\{n\Big(1-\frac{n-1}{N}\Big)-\frac{1}{2}\}],
\end{align}
\begin{align}
	G_1D_1^* &= \frac{1}{Z}\sum_m^M\sum_n^N G_1(m,n,t)D_1^*(m,n,t)\nonumber\\
	&\times \exp[-\frac{\hslash \omega_a}{K_B T}\{m\Big(1-\frac{m-1}{M}\Big)-\frac{1}{2}\}] \nonumber\\
	&\times \exp[-\frac{\hslash \omega_b}{K_B T}\{n\Big(1-\frac{n-1}{N}\Big)-\frac{1}{2}\}].
\end{align}
The rest of the elements, that is, lower diagonal elements in Eq. (\ref{choi-matrix}) are hermitian conjugates of the upper diagonal terms. 

\bibliographystyle{apsrev4-1}
\bibliography{reference}

\end{document}